\documentclass[a4paper,12pt]{article}

\pdfoutput=1
\usepackage{jheppub}

\def\sec#1{\S\ref{#1}}

\def\App#1{Appendix \ref{#1}}

\definecolor{rust}{rgb}{0.8,0.2,0.2}
\definecolor{green}{rgb}{0.1,0.8,0.2}
\title{Non-dissipative hydrodynamics: Effective actions versus entropy current}

\author[a]{Jyotirmoy Bhattacharya}
\author[b]{Sayantani Bhattacharyya}
\author[c]{Mukund Rangamani}
 \affiliation[a]{Kavli Institute for the Physics and Mathematics of the Universe (WPI),\\
 The University of Tokyo, Kashiwa, Chiba 277-8583, Japan.}
\affiliation[b]{Harish Chandra Research Institute,\\Chhatnag Rd, Jhusi, Allahabad, India.}
\affiliation[c]{ Centre for Particle Theory \& Department of Mathematical Sciences,\\
Science Laboratories, South Road, Durham DH1 3LE, UK.}

\emailAdd{jyotirmoy.bhattacharya@ipmu.jp}
\emailAdd{sayanta@hri.res.in}
\emailAdd{mukund.rangamani@durham.ac.uk}

\abstract{While conventional hydrodynamics incorporating dissipative effects is hard to derive from an action principle, it is nevertheless possible to construct classical actions when the dissipative terms are switched off. In this note we undertake a systematic exploration of such constructions from an effective field theory approach and argue for the existence of non-trivial second order non-dissipative hydrodynamics involving pure energy-momentum transport. We find these fluids to be  characterized by five second-order transport coefficients based on the effective action (a three parameter family is Weyl invariant). On the other hand since all flows of such fluids are non-dissipative, they entail zero entropy production; one can therefore understand them using the entropy current formalism which has provided much insight into hydrodynamic transport. An analysis of the most general stress tensor with zero entropy production however turns out to give a seven parameter family of non-dissipative hydrodynamics  (a four parameter sub-family being Weyl invariant). The non-dissipative fluids derived from the effective action approach are a special case of the fluid dynamics constrained by conservation of the entropy current. We speculate on the reasons for the mismatch and potential limitations of the effective action approach.  }

\begin{document}

\begin{flushright} \small{IPMU12-0196, DCPT-12/43} \end{flushright}

\maketitle

\flushbottom

%%%%%%%%%%%%%%%%%%%%%%%%%%%%%

%______________________________________________
\section{Introduction}
\label{sec:intro}
%______________________________________________

Fluid dynamics is an effective description, valid over spatio-temporal length scales large compared to the mean free path of any interacting system. Traditionally the variables of fluid dynamics are the local fluid velocity, temperature (or energy density) and the local charge densities. The dynamical equations are simply the conservation equations for the stress tensor and other conserved currents. The basic phenomenological input of fluid dynamics are the constitutive relations where the stress tensor and other conserved currents are expressed in terms of the fluid variables i.e., velocity, temperature and the charge densities.

But these constitutive relations cannot be completely arbitrary. Because fluid dynamics is always a corse-grained description of some underlying microscopic quantum theory, it must satisfy all the consistency requirements arising from the quantum dynamics of the fundamental degrees of freedom (in an appropriate long-wavelength fluid limit).

First of all the constitutive relations are constrained by the symmetries of the system. But more importantly we must use a local version of second law to further constrain the phenomenology of the stress tensor and currents. In particular, we demand that the entropy should locally increase on every time dependent solution of the fluid equations i.e., there should exist at least one entropy current whose divergence is always non-negative for every  fluid flow consistent with the equations of motion. It turns out that this particular condition constrains the possible form of the constitutive relations to a large extent, a fact that has been well-appreciated for a while, cf., \cite{landau}.

However, in spite of decades of work on fluid dynamics, we do not yet know if the aforementioned requirements suffice to guarantee the full consistency of the hydrodynamical equations, when we further demand that such a fluid is an infra-red effective description of some microscopic quantum theory. We do not yet have a formulation which can capture all these microscopic constraints simultaneously (nor have we established what the full set of constraints are). What we do know is that it appears to be sensible to use a working definition of hydrodynamics in terms of the symmetries and a local version of the second law, cf., \cite{Loganayagam:2008is,Romatschke:2009kr,Son:2009tf, Bhattacharya:2011tra, Loganayagam:2011mu,Bhattacharyya:2012nq} for recent developments which explore this philosophy in detail.

Generically fluid dynamics involves dissipation; to leading order in gradients this is reflected by the parabolic nature of the energy-momentum conservation $\nabla_\mu T^{\mu\nu} = 0$ arising from viscous terms. This has entailed that traditional analyses of fluid dynamics as an effective field theory involves working with the equations of motion. It is difficult to write some quantity like an `effective action' which upon extremization  produces these equations thereby ensuring all such consistencies at one go. 

It is however intriguing to ask if it is possible to consider purely non-dissipative fluids. These clearly have a chance of being derivable from an effective action, and provide an opportunity to explore the microscopic constraints on fluid dynamics. Clearly, ideal hydrodynamics (a perfect fluid) is an example of such non-dissipative dynamics. A-priori it is not clear if we can deform away from perfect fluidity by higher order terms; if this were not possible then we would be discussing a rather boring class of examples. Our program of using effective actions will be essentially re-deriving thermodynamics (which clearly can be done using an Euclidean path integral to compute the free energy). Note that since we are switching off dissipative terms, we are 
guaranteed to have an entropy current which is identically conserved.

Fortunately, it turns out that the space of non-dissipative fluid dynamical theories is more than ideal fluids (as we argue below). One can proceed either by starting with a dissipative fluid system, setting the dissipative transport terms to zero and checking that the resulting system satisfies known constraints of an autonomous hydrodynamics, viz., symmetry principles and entropy conservation. However, we can also proceed by trying to construct and effective action, since any action formalism as argued earlier could capture the physics of non-dissipative fluids. The main question is what is the organizing principle for such an action. It turns out the one can construct an effective action for hydrodynamics in $d$-dimensional spacetime  in terms of $d-1$ scalar fields $\phi^I$, which are the (Lagrangian) labels for local fluid element as a function of the fixed spacetime coordinates $x^\mu$. Alternately one can view the system in an Eulerian language by prescribing the spatial position as a function of the fluid element labels and time. The effective action is constructed as a re-parameterization invariant action respecting volume preserving diffeomorphisms in the $\phi$ space. This approach has been used to describe ideal fluids over the years motivated by an attempt to write action principles; cf.,   \cite{Taub:1954zz,Carter:1973fk,Carter:1987qr,Brown:1992kc,Leutwyler:1996er,Jackiw:2004nm,Dubovsky:2005xd, Nicolis:2011ey, Dubovsky:2011sj,Dubovsky:2011sk}.

The most modern analysis of such effective actions for hydrodynamics was carried out recently in  \cite{Dubovsky:2011sj}. In particular, they argued that such an effective field theory naturally implements entropy conservation as an identity (a consequence of the volume preserving diffeomorphisms) thus making them ideal for explorations of non-dissipative hydrodynamics. For both uncharged and charged fluids they have shown how their formulation produces ideal fluid dynamics and they discuss some higher order corrections to exemplify their approach.\footnote{This effective action formalism has been shown to reproduce certain features of anomalous hydrodynamic transport \cite{Dubovsky:2011sk} in two spacetime dimensions as well as describing features of parity violation in 3 spacetime dimensions \cite{Nicolis:2011ey}. We will comment on these constructions at appropriate stages in our discussion.} We will be interested in a general analysis of such classes of actions; we will classify all possible higher order corrections (to second order in an appropriate derivative counting scheme) and understanding the resulting conditions on the transport coefficients. 

Before proceeding however, we should first issue a disclaimer: while we will analyze the constraints arising on hydrodynamics from the existence of a classical effective action, we will not be able to argue that the resulting classical theory itself arises from an underlying unitary quantum dynamics upon coarse-graining. For much of the text we will analyze the classical description of non-dissipative hydrodynamics, relegating to \sec{sec:conclude} some speculations on whether one can realize such from well behaved quantum dynamics.

 The first task we undertake is to examine the effective action built from the Lagrangian fluid element variables systematically in a gradient expansion. We need at every order in the expansion to classify the number of on-shell inequivalent operators built out of the variables $\phi^I$. As mentioned earlier  \cite{Dubovsky:2011sj} analyze the leading order piece in such a derivative expansion (we explain our derivative counting more carefully below). The number of allowed structures grows with increasing  order, but it easy to enumerate the relevant operators at two orders beyond the leading one. Focussing specifically on uncharged fluids, 
we  find that the first non-zero correction to the action arises only at second order, where $5$ distinct on-shell inequivalent operators exist. As a result we can have 5 free {\em parameters}\footnote{In this note, by the word `parameter', we always mean an arbitrary function of temperature or entropy density. Thus our parameters always refer to transport functions.} (see Eq. \eqref{2ndaction}) entering in our effective action and therefore generates a stress tensor with 5 independent transport coefficients (see Eq. \eqref{stressstep}).
   
We then analyse how the existence of an entropy current with zero divergence constrains the possible form of the stress tensor by generalizing the analysis of \cite{Bhattacharyya:2012nq}. This reveals that one has a second order stress tensor with 7 independent transport coefficients cf., Eqs. \eqref{stresderi1}, \eqref{fconst1} below.

Therefore the two stress tensors, derived using these two methods almost agree except these two free transport coefficients. We find that the stress tensor determined from action is a special case of the stress tensor determined using the entropy current analysis, provided we do some identification of the parameters, cf., Eq. \eqref{identif}.

This is what we should expect since the `action formalism' always admits a conserved entropy current by construction. However it is interesting that from this action we do not get the most general stress tensor allowed by the existence of a conserved entropy current. This  can be interpreted either as an indication that the `entropy current technique' is not sufficient  or on the contrary as a hint that some further generalization of this `action formalism' is possible.

This note is organized as follows. In \sec{sec:form} we shall briefly discuss the effective action formulation of non-dissipative fluids following the treatment of \cite{Dubovsky:2011sj} and extend it to second order to derive the stress tensor. In \sec{sec:entropy} we shall specialize the entropy current analysis of \cite{ Bhattacharyya:2012nq} to the case of zero divergence entropy current and derive the constraints on the transport coefficients. 
In order to compare the two computations in \sec{sec:frame} we first briefly describe the frame invariant formulation which allows to compare fluid dynamical data derived in inequivalent frames. We then show that one can sensibly compare the two stress tensors (and entropy currents) derived in \sec{sec:form} and  \sec{sec:entropy} respectively and present the identification of the parameters appearing in the preceding two sections. In section \sec{sec:conclude} we shall conclude with some implications of our discussion and some open questions. We collect some useful intermediate results and conventions in \App{s:conventions} and present a brief analysis of parity violating fluids in 3 spacetime dimensions in \App{s:parityodd}.

%______________________________________________
\section{Effective actions for non-dissipative hydrodynamics}
\label{sec:form}
%______________________________________________
 
 As mentioned in the introduction, it should be possible to construct and effective action for non-dissipative hydrodynamics. Such a construction should satisfy two primary requirements:
\begin{itemize}
\item  The Euler-Lagrange equations arising from the effective action which lead to the equations of motion of the theory should have no more content than energy-momentum conservation. The latter arise as a consequence of diffeomorphism invariance of the action, since the stress-tensor can be obtained by varying the action with respect to the background metric.
\item Lack of dissipative effects means that in such a formalism one should be able to identify a conserved current, which we interpret as the entropy current $J_S^\mu$.
\end{itemize} 

The requirement of equations of motion having no more dynamical content than stress tensor conservation is a 
strong one. Usually while it is true that equations of motion imply $\nabla_\mu T^{\mu\nu}= 0$ the converse often fails to hold. 

The key point is identifying the correct degrees of freedom which achieves this; such a construction has been known for a long time based on what we might call, Lagangian fluid variables. Intuitively we want some symmetry in the field space that mimics the diffeomorphism invariance of the background spacetime. This can trivially be achieved by demanding that that configuration space of our classical action be parameterized by canonical field  variables which respect field redefinition invariance (which is the analog of diffeomorphism in field space). Then by passing to some gauge fixed version a la,, static gauge we can argue that field variations which lead to Euler-Lagrange equations can be conflated with background diffeomorphisms thereby ensuring that energy-momentum conservation being the dynamical equations of the theory; cf., \cite{Julia:1998ys} for observations relating to this point.

The fields of interest are labels for individual fluid elements viewed as a function of background spacetime coordinates.  Such formulations have been described for perfect fluids (both relativistic and non-relativistic) for a very long time as mentioned in the introduction. We will use the recent analysis of  \cite{Dubovsky:2011sj} who were able to use these variables to motivate an effective action for non-dissipative hydrodynamics. While \cite{Dubovsky:2011sj} discuss both neutral and charged fluids we will focus exclusively on neutral fluids in our discussion below.

%~~~~~~~~~~~~~~~~~~~~~~~~~~~~~~~~~~~~~~~~~~~~~~~
\subsection{The fundamental fields of hydrodynamics}
\label{s:ff}
%~~~~~~~~~~~~~~~~~~~~~~~~~~~~~~~~~~~~~~~~~~~~~~

Let us quickly review the ingredients in the construction of \cite{Dubovsky:2011sj} starting with uncharged fluids. As mentioned earlier we want to work with local fluid elements, and use fields $\phi^I$ describe the position of the local fluid element in space at an instant of time. We will work with $d$-dimensional fluids, so $i = 1, 2, \cdots , d-1$. We further fix  the geometry of the spacetime in which the fluid propagates and take coordinates $x^\mu$ to be an appropriate chart on this spacetime manifold, with metric $g_{\mu\nu}$.

Since we are trying to tag local fluid elements, we expect that the description of the low energy effective dynamics enjoys translational and rotational invariance of these co-moving coordinates. This tells us that the effective action should be constructed out of the derivatives of the fields $\phi^I$ and be suitably rotationally invariant. In addition, a local version of Liouville theorem in this configuration space of the $\phi^I$ demands invariance under arbitrary reparameterizations of the $\phi^I$, i.e., the Lagrangian should be invariant under 
\begin{equation}
\phi^I \to \xi^I(\phi) \ , \qquad \text{Jacobian}(\xi,\phi) = 1
\label{}
\end{equation}	
with the condition of the Jacobian being forced on us by the fact the volume of configuration space be unchanged. Clearly, such  a symmetry is generated by the diffeomorphisms of the $\{\phi^I\}$ space by vector fields that are divergence free, i.e., locally we need that $\phi^I \to \phi^I + \xi^I$ with ${\cal D}_I \, \xi^I  =0$. Here ${\cal D}_I$ denotes the covariant derivative in the manifold parameterized by $\{\phi^I\}$ which we call ${\cal M}_\phi$. The symmetry we demand is invariance under volume preserving diffeomorphisms of ${\cal M}_\phi$; this symmetry group is often denoted as $\text{Sdiff}({\cal M}_\phi)$.

Before proceeding to construct an action with the symmetries described above let us note one consequence of the field reparametetrization invariance. Consider the current one-from obtained by taking the spacetime dual of the volume form of ${\cal M}_\phi$:
\begin{equation}
J = \star \left(d\phi^1 \wedge d\phi^2 \wedge \cdots d\phi^{d-1} \right) ,
\label{}
\end{equation}	
or directly in components:
\begin{equation}
J^\mu = \frac{1}{(d-1)!} \, \epsilon^{\mu\alpha_1\cdots \alpha_{d-1}} \, \epsilon_{I_1\cdots I_{d-1}} \, \prod_{j=1}^{d-1}\; \partial_{\alpha_j} \phi^{I_j} \,.
\label{Jdef}
\end{equation}	
This current is is trivially conserved 
\begin{equation}
\nabla_\mu J^\mu = 0 \,.
\label{}
\end{equation}	
We will soon see that this current is to be viewed as the entropy current $J^\mu_S$ and the volume preserving diffeomorphism symmetry ensures this conservation automatically.

We will find it convenient to split the current $J^\mu$ into a scalar and a normalized $d$-vector:
\begin{equation}
J^\mu = s\, u^\mu \,, \qquad s = \sqrt{-J_\mu J^\mu} \,, \qquad u^\mu \, u_\mu = -1
\label{sdef}
\end{equation}	
and build the Lagrangian out of these fields (respecting spacetime diffeomorphism invariance). The notation is intentionally suggestive: $s$ is to be identified with the entropy density of the fluid and $u^\mu$ with its normalized $d$-velocity.

%~~~~~~~~~~~~~~~~~~~~~~~~~~~~~~~~~~~~~~~~~~~~~~~
\subsection{Operator dimensions: setting up the gradient expansion}
\label{s:}
%~~~~~~~~~~~~~~~~~~~~~~~~~~~~~~~~~~~~~~~~~~~~~~

Now we are left with the task of writing down the general effective action with the symmetries of the $\phi$ fields. To get off the ground, we need to understand the canonical operator dimensions for such a construction. Since the $\phi^I$ fields are to be viewed as Goldstone modes for the embedding of the fluid into the background spacetime, they turn out to have mass dimension $\left[\phi^I \right]  = -1$. This effectively implies that we treat them as phase fields and consequently $\left[d\phi^I \right] = 0$ implying that $J^\mu$ is a dimensionless operator in the field space (and a tensor density in physical spacetime).

Our task is now clear: we should first classify all operators ${\cal O}_\Delta$ with canonical scaling dimension $\Delta = 0, 1,2 , \ldots $ etc.. However, since $\left[d\phi^I\right]= 0$ even an operator with $\Delta = 0$ can have arbitrary many derivatives when we attempt to build it out of the fundamental fields $\phi^I$. Fortunately, the invariance under $\text{Sdiff}({\cal M}_\phi)$ comes to our rescue. We use this large symmetry to make the following assertion:

\paragraph{Claim:} The action $S$ is a functional of the entropy current $J^\mu$ and its derivatives, the latter being treated as the zeroth order operator. The full action can be written  in derivative expansion as an infinite sum
  $$ S = \sum_{i=0}^\infty S_i $$ 
 where $i^{\rm th}$ order the action $S_i$ contains $i$ space-time derivatives. In addition to the operator $J^\mu$ (equivalently $s$ and $u^\mu$) we should also keep track of terms that involve operators built out of the background geometry. In the present discussion of neutral fluids these are simply operators built out of the intrinsic geometry of the background, and hence involve the metric tensor $g_{\mu\nu}$ and the curvature tensors built from it.
    
We now proceed to analyze the effective action in detail at the first three orders in the spacetime derivative expansion, based on the above assertion without providing a definitive proof of it. Note that one really should classify all invariants under $\text{Sdiff}({\cal M}_\phi)$ at a given order in the derivative counting described above; it is the latter that ensures that the dynamics of $\phi^I$ is equivalent to energy-momentum conservation while guaranteeing entropy conservation.

%~~~~~~~~~~~~~~~~~~~~~~~~~~~~~~~~~~~~~~~~~~~~~~~
\subsection{Zeroth order: Ideal fluid hydrodynamics}
\label{s:}
%~~~~~~~~~~~~~~~~~~~~~~~~~~~~~~~~~~~~~~~~~~~~~~
  
  At zero derivative order the only scalar that  can be constructed out $\phi^I$ respecting $\text{Sdiff}({\cal M}_\phi)$ invariance is the norm of $J^\mu$ which according to \eqref{sdef} is proportional to the square of the entropy density. Therefore the zeroth order action can be written as (the normalization is to ensure a simple stress tensor)
  \begin{equation}\label{action0}
  \begin{split}
  S_0 =  -2 \, \int \sqrt{-g}\;  f(s)
  \end{split}
  \end{equation}
  where $f(s)$ is some arbitrary function of the entropy density. 
  
 Given this effective action, we can vary with respect to the background spacetime metric using \eqref{idex2} to obtain the  stress tensor\footnote{Henceforth for any function $F(s)$ we denote by $F'$ the derivative with respect to entropy density $s$ i.e.,  $F'(s) \equiv \frac{dF}{ds}$.} 
 \begin{equation}\label{stress0}
 \begin{split}
 T^{\mu\nu}_{(0)} &= \frac{1}{\sqrt{-g}}\frac{\delta S_0}{\delta g_{\mu\nu}}\\
 &= (s\, f'(s)-f(s) ) \, g^{\mu\nu} +s\, f'(s) \, u^\mu \, u^\nu \\
 & = \varepsilon\, u^\mu u^\nu  + p\, P^{\mu\nu}
 \end{split}
 \end{equation}
In the last line we have identified the energy density $\varepsilon(s) = f(s)$ and used thermodynamics to identify $p(s) = \left(s\, f'(s) - f(s)\right)$ with the  pressure. $P^{\mu\nu} \equiv g^{\mu\nu}  + u^\mu\, u^\nu$ is the projector\footnote{Our conventions for fluid dynamical tensors are collected in \eqref{notation}.}
 in the direction transverse to $u^\mu$. After this identification the stress tensor presented in \eqref{stress0} has taken the form of the ideal stress tensor for an uncharged fluid. Thus the Lagrangian formulation for an ideal fluid simply involves writing down the energy in terms of the entropy density. 
 
 Let us quickly check the equations of motion: extremizing this action with respect to the  $\phi^I$  we get:
 \begin{equation}\label{equation0}
 \begin{split}
&\left(\epsilon^{\nu\mu_1\mu_2 \cdots \mu_{d-1}} \, \prod_{i\neq 1} \; \partial_{\mu_i}\phi_{I_i} \right) \nabla_{\mu_1}[f'(s) u_\nu] =0\\
& \Rightarrow \;\;(u^\nu P^{\mu\alpha} - u^\alpha P^{\mu\nu}) \nabla_\alpha [f'(s) u_\nu] =0\\
 \end{split}
 \end{equation}
 In the second line we have used the fact that $J^\mu \nabla_\mu \phi_I = u^\mu \nabla_\mu \phi_I =0$.
 
There are thus $d-1$ Euler-Lagrange equations (one for each $\phi^I$) which we can conveniently assemble into the conservation equations of the energy-momentum tensor projected transverse to the velocity $u^\mu$.
The equations in \eqref{equation0} automatically follow from simply $P_{\mu\alpha}\, \nabla_\nu T^{\mu\nu}_{(0)} = 0$. Further, as already argued in  \cite{Dubovsky:2011sj} the remaining conservation equation $u^\mu \nabla_\nu T^{\mu\nu}_{(0)}= 0$ is implied by the conservation of entropy current $J^\mu$ and therefore identically zero. Therefore the equations of motion for the basic $\phi_I$ fields have no more content that that of the conservation of stress tensor as desired. This is of course in keeping with the general motivation for working with the fields $\phi^I$; the only novelty is that $\text{Sdiff}({\cal M}_\phi)$ invariance allows us to trade energy conservation for entropy conservation.

%~~~~~~~~~~~~~~~~~~~~~~~~~~~~~~~~~~~~~~~~~~~~~~
\subsection{First order: absence of viscous effects}
%~~~~~~~~~~~~~~~~~~~~~~~~~~~~~~~~~~~~~~~~~~~~~~

At first order in derivative expansion we can construct one non-trivial scalar out of $J^\mu$, namely $J^\mu\nabla_\mu s$.\footnote{The one other scalar built out of $J$ is its  divergence of $\nabla_\mu J^\mu$ which is identically vanishes.} Therefore naively we can write an action at first order 
\begin{equation}
S_1 = \int \sqrt{-g} \, f_1(s) \, J^\mu \nabla_\mu s 
 = \int \sqrt{-g} \, J^\mu \nabla_\mu {\tilde f}_1(s) 
\label{}
\end{equation}	
with for some $\frac{d{\tilde f}_1(s)}{ds} = f_1(s)$. This term is however a total derivative, for we can rewrite $S_1$ as
\begin{equation}
 S_1 = \int \sqrt{-g}\left[\nabla_\mu \left(\tilde f_1(s) \, J^\mu\right) -\tilde f_1(s) \; \nabla_\mu J^\mu \right] 
\label{}
\end{equation}	
with the second term being identically zero. 

In our analysis we shall ignore all the total derivative terms in the action, since they don't contribute to equations of motion. It then follows that $S_1$ can be ignored in what follows; therefore the first order correction to the stress tensor is also zero. The first non-trivial correction appears at second order.\footnote{This statement is not true in three dimensional fluids with parity violation: we can construct a one-derivative operator $h_1(s) \, \epsilon^{\mu\nu\rho} J_\mu\, \nabla_\nu J_\rho$ which is not a total derivative. This contributes to transport; see \cite{Nicolis:2011ey} and also \App{s:parityodd} for details.}

This is of course intuitive; we are describing a non-dissipative fluid with an exactly conserved entropy current. At first order we know that we can have viscous contributions to the stress tensor, but these generically produce entropy, since $\nabla_\mu  J_S^\mu \propto \eta \, \sigma_{\mu\nu} \, \sigma^{\mu\nu} + \zeta\, {\Theta}^2$ where $\sigma_{\mu\nu}$ and $\Theta$ are the fluid shear and expansion respectively, cf., \eqref{notation}. Requiring entropy conservation for arbitrary configurations forces upon us $\eta = \zeta = 0$. 

This point is likely to make the reader uncomfortable; while we know that fluids with $\zeta =0$ are physical (e.g., conformal fluids), vanishing of $\eta$ seems a bit strange. We postpone the discussion of whether this is physically acceptable to \sec{sec:conclude}.

%~~~~~~~~~~~~~~~~~~~~~~~~~~~~~~~~~~~~~~~~~~~~~~
\subsection{Second order: novel non-dissipative fluids}
%~~~~~~~~~~~~~~~~~~~~~~~~~~~~~~~~~~~~~~~~~~~~~~

We now turn to the second order in gradients and argue that we have a five distinct on-shell inequivalent operators that we can write down in the effective action. First, note that now  we encounter two types of scalar operators: ones that involve explicit curvature tensor and the others that do not.

There are two curvature dependent scalars:
\begin{equation}
R \, \qquad \text{and} \qquad R_{\mu\nu}\, u^\mu\, u^\nu \,, 
\label{cbasis2}
\end{equation}	
which can enter our effective action multiplied by arbitrary functions of $s$. 

The scalars without curvature can again be divided into two subclasses,
ones where both the derivatives are acting on the same $J^\mu$ (e.g. $J^\mu\nabla^2J_\mu$) and the others where the two derivatives are acting on two different $J^\mu$ (e.g. $\nabla_\mu J_\nu \nabla^\mu J^\nu$). But any term which is of the first type can always be recast into a term of second type upto a total derivative. Therefore for the second order action we only need to classify those  scalars which are constructed as product of terms with single derivative acting $J^\mu$.

Equivalently we can view the construction in terms of $s$ and $u^\mu$ and ask for scalars that are built out of the one-derivative vector $\nabla_\mu s$ and the one-derivative two tensor $\nabla_\mu u_\nu$. A-priori we would write down six such scalars:
\begin{align}
& \left(u^\mu\nabla_\mu s\right)^2, &   
\left(u^\alpha\, \nabla_\alpha u^\mu\right)^2,   & &
 \left(P^{\mu\nu}\nabla_\nu s\right)^2,
\nonumber \\
&\left(u^\alpha\, \nabla_\alpha u_\mu\right) \nabla^\mu s , & 
(\nabla_\mu u_\nu)(\nabla^\mu u^\nu), & &
(\nabla_\mu u_\nu)(\nabla^\nu u^\mu)
\label{ncbasis2}
\end{align}	
where for any vector $A^\mu$, $(A^\mu)^2$ denotes $(A_\mu A^\mu)$ for brevity. We now argue that in fact  from the basis of 8 operators contained in \eqref{cbasis2} and \eqref{ncbasis2} we only need to consider 5 operators.

Generically the action constructed at second order will have a field redefinition ambiguity. In order to fix this we can work only with those operators which are 
on-shell  inequivalent. The equations of motion arising from the zeroth order action $S_0$ \eqref{action0}, together with $\nabla_\mu J^\mu = 0$ allows us to relate derivatives acting on $s$ to derivatives acting on $u^\mu$. So we only need to consider two-derivative scalars obtained from contractions the two-tensor $\nabla_\mu u_\nu$. Using the decomposition of this two tensor into a transverse vector (acceleration), symmetric traceless \& anti-symmetric two-tensors (shear \& vorticity) and a scalar (expansion) (cf., \eqref{notation}) i.e.,
\begin{equation}
\nabla_\mu u_\nu = -u_\mu\, {\mathfrak a}_\nu + \sigma_{\mu\nu} + \omega_{\mu\nu} + \frac{\Theta}{d-1}\, P_{\mu\nu}\,,
\label{pudecomp}
\end{equation}	
we find a basis of four scalar operators:
\begin{equation}
{\mathfrak a}^2\,, \quad \sigma^2 \,, \quad \omega^2 \,, \quad \Theta^2 \,. 
\label{intbasis2}
\end{equation}	
Furthermore, using the the fact that $R_{\mu \nu} u^\mu u^\nu$ is given by a commutator of derivatives acting on $u^\mu \, u^\nu$, we can eliminate it in favor of operators built from the velocity \eqref{intbasis2}, see \App{s:conventions} for details. This is physically the most intuitive basis given the observations earlier about $\eta = \zeta = 0$.  Hence the action at second order finally has five terms.

From a computational perspective however, we find it convenient to switch to a different basis which is more efficient for variation of the action with respect to the metric. Introduce thus the second order action:
\begin{equation}
\label{2ndaction}
\begin{split}
  S_2 =&~  \int \sqrt{-g} \bigg[ K_1(s) \; P^{\alpha\beta}\, (\nabla_\alpha u_\nu)\, (\nabla_\beta u^\nu)
 + K_2(s) \; (\nabla_\mu u_\nu)\, (\nabla^\nu u^\mu) \\
&~~~~~~~~~~~~~+ K_3(s) \; (J^\mu\nabla_\mu s)^2
 + K_4(s) \; \nabla_\mu s\,\nabla^\mu s  + K_5(s) \; R\bigg]
 \end{split}
\end{equation}
which satisfies all our physical requirements.

Varying this action with respect to the metric using \eqref{idex2} we get stress tensor contribution:
\begin{equation}\label{2ndstress}
 \begin{split}
  &\Pi^{\mu\nu} = \frac{1}{\sqrt{-g}}\; \frac{\delta S_2}{\delta g_{\mu\nu}} 
= \sum_{i=1}^5 \; \Pi^{\mu\nu}_{(i)}\\
\end{split}
\end{equation}
where
\begin{subequations}
\label{stressstep}
\begin{align}
\Pi_{(1)}^{\mu\nu} =&~\left[\left(\frac{K_1 - s K_1'}{2}\right) 
P^{\alpha\beta}(\nabla_\alpha u_\theta)(\nabla_\beta u^\theta)\right]P^{\mu\nu}
+\left[\frac{K_1}{2}P^{\alpha\beta}(\nabla_\alpha u_\theta)(\nabla_\beta u^\theta)\right]
u^\mu u^\nu\nonumber \\
&~+K_1 {\mathfrak a}^2 u^\mu u^\nu - K_1 (\nabla^\mu u_\alpha)(\nabla^\nu u^\alpha)
 + K_1 P^{\alpha\beta}(\nabla_\alpha u^\mu )(\nabla_\beta u^\nu)\nonumber \\
&~-\nabla_\alpha\left[u^{(\mu} P^{\alpha\beta}K_1 \nabla_\beta u^{\nu)}\right] 
- \nabla_\alpha\left[u^\alpha \,P^{\beta(\mu} K_1 \nabla_\beta u^{\nu)}\right]
+ \nabla_\alpha\left[K_1 u^{(\mu} P^{\nu)\beta}\nabla_\beta u^\alpha\right]  
\label{stena}
\end{align}
\begin{align}
\Pi_{(2)}^{\mu\nu} =&~\left[\left(\frac{K_2 - s K_2'}{2}\right) 
(\nabla_\alpha u_\beta)(\nabla^\beta u^\alpha)\right]P^{\mu\nu}
-\left[\frac{K_2}{2}(\nabla_\alpha u_\beta)(\nabla^\beta u^\alpha)
+ u^\alpha\nabla^{\beta}\left(K_2\nabla_\alpha u_\beta\right)\right]u^\mu u^\nu\nonumber \\
&~-\nabla_\alpha\left[K_2 u^{(\nu} \nabla^{\mu)} u^\alpha\right] 
- \nabla_\alpha\left[K_2 u^\alpha \nabla^{(\mu} u^{\nu)}\right] 
+ \nabla_\alpha\left[K_2 u^{(\nu} \nabla^\alpha u^{\mu)}\right]  
\label{stenb}
\end{align}
\begin{align}
\Pi_{(3)}^{\mu\nu}=&~\left[ \left(\frac{3 s^4 K_3 + s^5 K_3'}{2}\right) \Theta^2 
-s^4 K_3 \, u^\alpha\nabla_\alpha\Theta \right]P^{\mu\nu}
+ \left[\left(\frac{s^4 K_3}{2} \right)\Theta^2\right]u^\mu u^\nu 
\end{align}
\begin{align}
\Pi_{(4)}^{\mu\nu} =&~\left[ \left(\frac{K_4 + s K_4'}{2}\right) (\nabla s)^2 
+s K_4 \nabla^2 s\right]P^{\mu\nu}
- \left[\left(\frac{ K_4}{2} \right)(\nabla s)^2 \right]u^\mu u^\nu\ -K_4 \nabla^\mu s\nabla^\nu s  
\end{align}
\begin{align}
\Pi_{(5)}^{\mu\nu} =&~R\left[ \left(\frac{K_5 - s K_5'}{2}\right) 
P^{\mu\nu}
- \left(\frac{ K_5}{2} \right)u^\mu u^\nu\right]\ 
+(\nabla^\mu\nabla^\nu K_5- g^{\mu\nu}\nabla^2K_5) - K_5 R^{\mu\nu}
\end{align}
\end{subequations}
We note that some of the expressions above can be simplified upon using the standard decomposition of the covariant derivative of the velocity field $u^\mu$; see \App{s:conventions}. We have also used the entropy conservation equation in $\Pi^{\mu\nu}_{(3)}$ to express $J^\mu\,\nabla_\mu s = -s^2\,\Theta$.

That total stress tensor $T^{\mu\nu} = T_{(0)}^{\mu\nu} + \Pi^{\mu\nu}$ is of course conserved and by our earlier arguments this conservation implies the Euler-Lagrange equations for the action $S_0 + S_2$. Thus using this effective action approach we have constructed a five parameter (in the sense described in footnote 1) family of non-dissipative fluids.

%~~~~~~~~~~~~~~~~~~~~~~~~~~~~~~~~~~~~~~~~~~~~~~~
\section{Constraints arising from a conserved entropy current}
\label{sec:entropy}
%~~~~~~~~~~~~~~~~~~~~~~~~~~~~~~~~~~~~~~~~~~~~~~

We now turn to a different approach to describing non-dissipative fluid dynamics based on an analysis of the entropy current. This approach has been used to investigate neutral fluids initially in \cite{Romatschke:2009kr} and more recently \cite{Bhattacharyya:2012nq} has analyzed the constraints on regular dissipative fluids exhaustively up to second order. In this section we shall see how the existence of an exactly conserved entropy current constrains the possible form of the stress tensor at first and second order in a gradient expansion; here we shall follow analysis of \cite{Bhattacharyya:2012nq}.\footnote{While the analysis described here can be generalised to arbitrary dimensions, we will restrict attention to $d=4$ in this section mainly to avoid accidental relations (especially in $d=2,3$) which affect our choice independent tensor structures.}

To begin with we have to use symmetry and on-shell equivalence to have a naive count of the number of possible terms in a hydrodynamical stress tensor at first and second order in gradients. This has been done in \cite{Bhattacharyya:2012nq}; it turns out that the stress tensor at first order has 2 transport coefficients and at second order has 15 transport coefficients. Writing $T^{\mu\nu} = T_{(0)}^{\mu\nu} + \Pi^{\mu\nu}$ the  most general form of $\Pi^{\mu\nu}$ up to second order in gradients (constrained only by symmetry and on-shell equivalence) is given by the following expression:
 \begin{equation}\label{stresderi1}
\begin{split}
\Pi_{\mu\nu} =~&-\eta\, \sigma_{\mu\nu} - \zeta \, P_{\mu\nu} \Theta\\
~&+\;T\bigg[ \tau \, u^\alpha \nabla_\alpha \sigma_{\langle\mu\nu\rangle} + \kappa_1 \, R_{\langle \mu\nu\rangle} + \kappa_2 \, F_{\langle \mu\nu\rangle} +\lambda_0\,  \Theta\, \sigma_{\mu\nu}\\
&\qquad +\lambda_1\, {\sigma_{\langle \mu}}^\alpha\, \sigma_{\alpha\nu\rangle}+ \lambda_2\,  {\sigma_{\langle \mu}}^\alpha \, \omega_{\alpha\nu\rangle}+ \lambda_3\, {\omega_{\langle \mu}}^\alpha\, \omega_{\alpha\nu\rangle} + \lambda_4\, {\mathfrak a}_{\langle\mu}{\mathfrak a}_{\nu\rangle}\bigg]\\
&+T\, P_{\mu\nu}\bigg[\zeta_1\, u^\alpha\nabla_\alpha \Theta + \zeta_2 \, R + \zeta_3\, R_{00}
 + \xi_1 \, \Theta^2 + \xi_2\,  \sigma^2+ \xi_3 \, \omega^2 
+\xi_4 \, {\mathfrak a}^2 \bigg]
\end{split}
\end{equation}
with $T$ being the (local) temperature.  In writing this we have introduced some notation which are defined as follows in $d=4$ space-time dimensions:
\begin{equation}\label{notation}
\begin{split}
&u^\mu =\text{The normalised four velocity of the fluid}\\
&P^{\mu\nu} = g^{\mu\nu} + u^\mu u^\nu =\text{Projector perpendicular to $u^\mu$}\\
&\Theta = \nabla_\alpha u^\alpha = \text{Expansion}, \qquad
{\mathfrak a}_\mu = u^\alpha\,\nabla_\alpha u_\mu = \text{Acceleration}\\
&\sigma^{\mu\nu} = 
P^{\mu\alpha} P^{\nu\beta}\left(\frac{\nabla_\alpha u_\beta + \nabla_\beta u_\alpha}{2}
 - \frac{\Theta}{d-1}g_{\alpha_\beta}\right) = \text{Shear tensor}\\
&\omega^{\mu\nu} = 
P^{\mu\alpha} P^{\nu\beta}\left(\frac{\nabla_\alpha u_\beta 
- \nabla_\beta u_\alpha}{2}\right)=\text{Vorticity}\\
&F^{\mu\nu} = R^{\mu \alpha \nu \beta}\, u_\alpha u_\beta, \qquad R^{\mu\nu} 
= R^{\alpha\mu \beta\nu}g_{\alpha \beta}\,, \qquad R_{00} = R^{\mu\nu} \,u_\mu\,u_\nu \\
&\sigma^2 = \sigma_{\mu\nu}\sigma^{\mu\nu},~~~~\omega^2 = \omega_{\mu\nu}\omega^{\nu\mu}
\end{split}
\end{equation}
with $R_{\alpha\beta\gamma\delta}$ being the Riemann tensor of the background geometry.
Furthermore, we define a projection of any tensor $A_{\mu\nu}$ onto its symmetric transverse (to $u^\mu$) traceless part via 
\begin{equation}
A_{\langle\mu\nu\rangle} \equiv P_\mu^\alpha P_\nu^\beta\left(\frac{A_{\alpha\beta} + A_{\beta\alpha}}{2} - \left[\frac{A_{ab}P^{ab}}{d-1}\right]g_{\alpha\beta}\right) \,.
\label{}
\end{equation}	

A conserved entropy current is the special case of the entropy current with non-negative divergence. The constraints on the second order transport coefficients due to the existence of a positive divergence entropy current have been determined in \cite{Bhattacharyya:2012nq}.The calculation in this section involves  a slight modification of such a general analysis (which is of course valid for all fluids). In fact the analysis up to Section 4 of \cite{Bhattacharyya:2012nq} can be adapted unchanged. Subsequent discussions will differ, but in a very simple way, so that the results can easily be read off from the equations presented in Section 5.1 and Appendices A and B of \cite{Bhattacharyya:2012nq}.

Let us first briefly recall the logic  used in \cite{Bhattacharyya:2012nq} to determine the constraints on transport.  We shall then be able to see how to modify it to avoid dissipation and obtain the constraints on the transport coefficients. The upshot of our discussion will be that we will show that of the $\{2+15\}$ transport coefficients appearing in \eqref{stresderi1} at first and second order respectively, one fixes $\{2+13\}$ in terms of the five arbitrary functions appearing in the entropy current.\footnote{We will denote the number of arbitrary functions appearing in various quantities of interest at first and second order in gradients by $\{\#_\text{1$^{st}$ order} +  \#_\text{2$^{nd}$ order} \}$ for simplicity.}

Entropy current has two parts, one is a canonical piece and the other is the correction. The canonical piece is completely fixed in terms of the zeroth order entropy current and $\Pi^{\mu\nu}$, the gradient correction to the ideal stress tensor.
\begin{equation}
J^\mu = J^\mu_{can} + J^\mu_{cor}\,, \qquad J^{\mu}_{can} = s\, u^\mu -\frac{u_\nu\, \Pi^{\mu\nu}}{T}
\label{ecurrentG}
\end{equation}	
In \cite{Bhattacharyya:2012nq} a particular fluid frame (Landau frame) was been chosen by demanding 
\begin{equation}
 u_\nu \, \Pi^{\mu\nu} =0
\label{landau}
\end{equation}	
In this frame $J^\mu_{can}$ is equal to entropy density times the local velocity.

The correction to the zeroth order canonical entropy current is determined using the fact that its divergence should never go negative on those fluid flows which satisfy the equations of motion. This argument remains unaffected even if we demand the total divergence of $J^\mu$ to be zero.\footnote{A-priori the most general entropy current for parity invariant fluids up to second order in gradients has $\{2+13\}$ parameters. By examining the number of independent scalars produced when one considers $\nabla_\mu J^\mu$ we find that $\{2+6\}$ of these have to be set to zero leaving behind the $\{0+7\}$ parameter set quoted in \eqref{entcurgen}.} Therefore we have the same $J^\mu_{cor}$ as in Eq.(1.4) of \cite{Bhattacharyya:2012nq}. The final form of $J^\mu$ in Landau gauge (correct up to second order) which we should consider is given by:  
\begin{equation}\label{entcurgen}
 \begin{split}
  J^\mu=&~ s \,u^\mu + \nabla_\nu\left[2\, A_1 \, u^{[\mu} \, \nabla^{\nu]} T\right] 
+ \nabla_\nu (A_2 \,T \,\omega^{\mu\nu})\\
&~+ A_3\left(R^{\mu\nu}-\frac{1}{2} g^{\mu\nu}R\right)u_\nu
 + \left(\frac{A_3}{T} + \frac{dA_3}{dT}\right)\left[\Theta\, \nabla^\mu T 
-P^{\alpha \beta} \nabla_\beta u^\mu\, \nabla_\alpha T \right]\\
&~+(B_1 \, \omega^2 + B_2 \, \Theta^2 + B_3\, \sigma^2)\, u^\mu +
 B_4\left[\nabla_\alpha s\, \nabla^\alpha s \, u^\mu + 2 \,s\, \Theta\, \nabla^\mu s\right]\\
 \end{split}
\end{equation}

Given such an entropy current one then computes its divergence. The latter also is best viewed in the decomposition of the canonic contribution and the correction piece. Using the equations of motion for $T_{(0)}^{\mu\nu} + \Pi^{\mu\nu}$  together with thermodynamic relations one finds (in $d=4$):
  \begin{equation}\label{prev1}
  \nabla_\mu J^\mu = \nabla_\mu J^\mu_{can} + \nabla_\mu J^\mu_{cor} =-\frac{1}{T} \left(\sigma_{\mu\nu}\, \Pi^{\mu\nu} + \frac{\Theta}{3}\,  P_{\mu\nu}\, \Pi^{\mu\nu} \right) + \nabla_\mu J^\mu_{cor}
 \end{equation}
 The key point to note is that the divergence of the canonical piece always involves $\Pi^{\mu\nu}$ as one of the factor and therefore will contain the information about the transport coefficients. 

For generic fluids in \cite{Bhattacharyya:2012nq} the idea was to rewrite \eqref{prev1} as a sum of perfect squares (up to fourth order in gradients) so as to ensure positivity. All the terms which could not be cast into the perfect square form were set to zero -- this gave the final constraints on the transport coefficients.

However, here we are interested in a zero divergence entropy current; this is the place where the present computation  differs from that of \cite{Bhattacharyya:2012nq}. We now proceed to outline the consequences of demanding that 
$\nabla_\mu J^\mu =0$.

Note that the two terms in \eqref{entcurgen} with coefficients $A_1$ and $A_2$ are explicitly of zero divergence and drop  out of our analysis. The rest of the five terms in $J^\mu_{cor}$ with five independent coefficients have non-zero divergence and therefore have to be cancelled against corresponding pieces coming from the divergence of $J^\mu_{can}$ and  constrain $\Pi^{\mu\nu}$.

When we compute the divergence of the entropy current given in \eqref{entcurgen} up to third order in gradients and  express the final answer in terms of the on-shell inequivalent data, each term turns out to have either $\sigma_{\mu\nu}$ or $\Theta$ as a factor. One can thus argue that schematically (please consult Appendix B of \cite{Bhattacharyya:2012nq} for explicit expressions):
\begin{equation}
\nabla_\mu J_{cor}^\mu =  \sigma_{\mu\nu} \, B^{\mu\nu} +\Theta  \, B
\label{}
\end{equation}	

The final expression of the divergence of the full entropy current is given by the following.
\begin{equation}\label{schemdiv}
\nabla_\mu J^\mu = - \sigma_{\mu\nu}\left(\frac{\Pi^{\mu\nu}}{T} - B^{\mu\nu}\right) - \Theta \left(\frac{\Pi^{\mu\nu}P_{\mu\nu}}{3\, T} - B\right)
\end{equation}
Generically the RHS of \eqref{schemdiv} will be zero if we set
\begin{equation}\label{schmconst}
\Pi^{\langle\mu\nu\rangle} = T\, B ^{\langle\mu\nu\rangle}+ Q^{\mu\nu}~~\text{and} ~~\Pi^{\mu\nu}P_{\mu\nu} = 3\,T\, B 
\end{equation}
where $Q^{\mu\nu}$ is any symmetric traceless tensor satisfying
$$\sigma_{\mu\nu} Q^{\mu\nu} = 0$$
 
 Using the schematic equations \eqref{prev1}, \eqref{schemdiv} and \eqref{schmconst} we can draw a few immediate conclusions.
 \begin{itemize}
 \item From \eqref{schmconst} upon demanding zero-divergence to third order, naively it appears that  we can determine the traceless part and the trace part of the stress tensor separately in terms of the parameters appearing in the entropy current.   This is generically true but with an important exception: when both the traceless and the trace part of the stress tensor contribute the same term to the divergence we can only fix a linear combination. This occurs for the terms $\lambda_0\, \Theta\, \sigma_{\mu\nu}$  in the traceless part and $\xi_2 \,\sigma^2$ in the trace part of our stress tensor \eqref{stresderi1}. Both of these end up contributing a term proportional to $\sigma^2 \, \Theta$ in the divergence of the canonical entropy current which is to be cancelled by the divergence of the `correction' part; this is manifest by writing:
\begin{equation}
\lambda_0 \,   \Theta \, \sigma_{\mu \nu}  + \xi_2 \, P_{\mu \nu}\,  \sigma^2 
= \frac{\lambda_0 + \xi_2}{2}\, \left( \Theta \, \sigma_{\mu \nu} +  P_{\mu \nu} \, \sigma^2 \right)    + \frac{\lambda_0 - \xi_2}{2} \left( \Theta \, \sigma_{\mu \nu} -  P_{\mu \nu} \, \sigma^2 \right)   
\label{}
\end{equation}	
Therefore our analysis will only be able to constrain $\lambda_0 + \xi_2$.
 
 \item Since $J^\mu_{cor}$ does not contain any term which is of first order in derivatives, both the tensor $B^{\mu\nu}$ and the scalar $B$  contain exactly two space-time derivatives. It then immediately follows from \eqref{schmconst}  that the first order correction to the stress tensor has to be zero, i.e., both shear viscosity $\eta$ and bulk viscosity $\zeta$ have to be zero, as is physically sensible for a non-dissipative fluid.

\item For parity invariant fluids  $Q_{\mu\nu}$ is proportional to ${\sigma_{\langle \mu}}^\alpha\,\omega_{\alpha\nu\rangle}$ at second order. Hence from \eqref{schmconst} it also follows that the transport coefficient $\lambda_2$, multiplying this term in $\Pi^{\mu\nu}$, will remain unconstrained by the entropy current analysis.
 \end{itemize}

We therefore can immediately conclude that $\{2+13\}$ transport coefficients appearing in \eqref{stresderi1} are fixed in terms of $\{0+5\}$ non-trivial arbitrary parameters appearing in $J^\mu_{cor}$. The explicit relations  are given by:
\begin{subequations}
\label{fconst1}
\begin{align}
&\eta = \zeta =0 \,,  \\
%\end{align}
%\begin{align}
& \tau = T\, \frac{dB_5}{dT} + 2 B_3 \,, \qquad 
\kappa_1  = A_3, \qquad 
\kappa_2= T\,\frac{dB_5}{dT} \,,  \\ 
%\end{align}
%\begin{align}
& \lambda_0 + \xi_2 = \left(B_3 - s \,\frac{dB_3}{ds}\right) - 2\,s\,T\left(\frac{ds}{dT}\right)B_4 \,,\\
%\end{align}
%\begin{align}
& \lambda_1 = T\,\frac{dB_5}{dT} \,,\qquad 
\lambda_3 = T \,\frac{dB_5}{dT} - 4 \,B_1 \,,
\label{lam13sol}
\\
%\end{align}
%\begin{align}
& \lambda_4=-\left[T^2\,\frac{d^2B_5}{dT^2}
 + T\,\frac{dB_5}{dT} + 2\,B_4 \,T^2\left(\frac{ds}{dT}\right)^2\right] ,
\end{align}
\begin{align}
&\zeta_1 = 2\,s\left(\frac{dB_5}{ds}\right) - \frac{2\,T}{3}\left(\frac{dB_5}{dT}\right) 
+ 2\, B_2 + 2\, B_4 s^2 - 2\, B_4 sT\left(\frac{ds}{dT}\right) ,
\nonumber \\
&\zeta_2 =\frac{1}{2}\left[s\,\frac{dA_3}{ds} - \frac{A_3}{3}\right] ,
\nonumber \\
 &\zeta_3 = s\, \frac{dA_3}{ds} + \frac{A_3}{3}
-\frac{2\,T}{3}\frac{dB_5}{dT} - 2\, B_4 \,T\,s\,\frac{ds}{dT}  \,,
\end{align}
\begin{align}
&\xi_1 =-s^2\, \frac{d^2B_5}{ds^2} - \frac{2\, T}{9}\left(\frac{dB_5}{dT} \right)
+ \left(B_2 - s \frac{dB_2}{ds}\right)
-\left[s^3\, \frac{dB_4}{ds} + s^2 \, B_4 + \frac{2\,s\,T}{3}\left(\frac{ds}{dT}\right)B_4\right] ,
\nonumber \\
&\xi_3 =-2\, B_4 \,T\,s\, \frac{ds}{dT} 
+ T \,\frac{dB_5}{dT}\left[\frac{s}{T}\,\frac{dT}{ds} -\frac{2}{3}\right] 
-s \,\frac{dB_1}{ds} + B_1\left[\frac{2\,s}{T}\,\frac{dT}{ds} -\frac{1}{3}\right] ,
\nonumber \\
 &\xi_4 = T^2\, s\, \frac{ds}{dT}\, \frac{dB_4}{dT}  
+ B_4\left[\frac{T^2}{3}\left(\frac{ds}{dT}\right)^2 + 4 \,T\, s \frac{ds}{dT}
 +2\, T^2\,s\,\frac{d^2s}{dT^2}\right]+ \frac{2}{3}\left(T\,\frac{dB_5}{dT} + T^2\,\frac{d^2B_5}{dT^2}\right) ,
\end{align}
\end{subequations}
where $ \frac{dB_5}{dT} = \frac{A_3}{T} + \frac{dA_3}{dT}$. 
This completes our analysis of non-dissipative fluids using the entropy current. To summarise have found a $\{0+7\}$ parameter family of such fluids; 5 parameters appear naturally in our parameterisation of the entropy current and two of the parameters that appear explicitly in the stress tensor viz., $\lambda_2$ and a linear combination of $\lambda_0, \xi_2$. We note in passing that $\lambda_1 = \kappa_2$ (which in turn in related to $\kappa_1$) for absence of entropy production.  
  
So far our discussion has been restricted to generic neutral fluids. One can further constrain the class of hydrodynamic theories by demanding that the fluid be conformal. Since the trace of the energy momentum tensor has to vanish for conformal fluids, this in particular implies that the only non-trivial transport coefficients are (cf., \cite{Baier:2007ix,Bhattacharyya:2012nq}): 
\begin{equation}
\tau = 3\, \lambda_0\,, \qquad \kappa_2 = 2\, \kappa_1 = \kappa \,, \qquad \lambda_1 \,,\qquad \lambda_2\,,\qquad \lambda_3
\label{}
\end{equation}	
All other second-order transport coefficients must vanish. A-priori since we have the remaining ten transport parameters determined in terms of five arbitrary functions one might suspect that there is no non-trivial solution. Working things out however we find that choosing\footnote{These relations can be read off directly from the analysis in Section 6 of \cite{Bhattacharyya:2012nq}, cf., Eq (6.7) of that paper. Noting that scale invariance also fixes $A_1 = 2\ a_3\,T$ and $A_2 =a_2\, T$, the only difference from the analysis of \cite{Bhattacharyya:2012nq} is the constraint on $B_2$: we find here that $B_2 = -\frac{2}{9}\, a_3\,T$, which differs by a factor of $-2$ from the value found earlier (for comparison please note that $a_1^\text{there} = 2\, a_3^\text{here}$). } 
\begin{equation}
A_3  = a_3\, T \,,\qquad B_1 = b_1\, T \,,\qquad  B_3 = b_3\, T \qquad \lambda_2 = \ell_2 \, T \,,
\label{ABpars}
\end{equation}	
ensures that all the conditions from $T_\mu^\mu =0$ are satisfied. The rest of the parameters $B_2$ and $B_4$ are determined in terms of $a_3$, in particular, $B_4 = -\frac{a_3}{9\, T^5}$ and $B_2 = -\frac{2}{9}\, a_3\, T$.
Note that scale invariance dictates that $s \propto T^3$. Changing basis of variables, the non-trivial transport coefficients can be taken to be $\tau$, $\lambda_1$, $\lambda_2$ and $\lambda_3$ (since $\kappa$ is given by $\lambda_1$ by \eqref{fconst1}).  We thus claim that the entropy analysis leads us to conclude the existence of a {\em four parameter family of conformal fluids}.\footnote{Since scale invariance fixes the temperature dependence of our transport coefficients, we really have only four parameters for conformal fluids.}

 %~~~~~~~~~~~~~~~~~~~~~~~~~~~~~~~~~~~~~~~~~~~~~~~
  \section{Frame invariant formulation and comparison}
  \label{sec:frame}
%~~~~~~~~~~~~~~~~~~~~~~~~~~~~~~~~~~~~~~~~~~~~~~

We  now turn to a comparison of the analysis of the preceding two sections of non-dissipative fluids. Whereas in   \sec{sec:entropy} we examined the constraints arising form the existence of the most general zero divergence entropy current, the analysis of \sec{sec:form} was predicated upon the existence of such a conserved entropy current. Therefore the stress tensor derived in  \sec{sec:form} must be a special case of the stress tensor derived in \sec{sec:entropy} up to a identification of parameters (we have conveniently recast the effective action formalism in terms of the fluid variables already); the same must also hold for the entropy current.
  
However, the analysis in the previous section has been done in Landau frame \eqref{landau} which defines the velocity $u^\mu$ (and temperature). Unfortunately the stress tensor obtained in \sec{sec:form} by varying the second order correction to the action is not in a Landau frame (this can be explicitly checked using  the explicit form of $\Pi^{\mu\nu}$ given in \eqref{2ndstress}). In fact, we can view the analysis of \sec{sec:form} to be in the `entropy frame' defined by the condition that entropy current does not receive any gradient correction, i.e., 
\begin{equation}
J^\mu = s \,u^\mu\,, \qquad \text{at all orders}
\label{entropyframe}
\end{equation}	
Given these observations one should not directly compare the $\Pi^{\mu\nu}$ given in \eqref{2ndstress} with the expression derived in \sec{sec:entropy}. We have to perform a field redefinition, i.e., redefine the velocity and temperature 
of one of the expressions so as to bring both answers into a common frame.
 
Happily, one can avoid doing an explicit field redefinition by choosing to compare only those combinations of $\Pi^{\mu\nu}$ which are invariant under such redefinitions of velocity and temperature (i.e., a frame transformation). Moreover the frame invariant combinations for both charged and uncharged fluids and superfluids have been worked out in \cite{Bhattacharya:2011tra}. Their formulation is applicable at the first non-trivial order of corrections to the stress tensors (and currents) in gradient expansion. Generically it is applied to the first order piece of $\Pi^{\mu\nu}$ as we encounter viscous effects at that order. But we have already seen that for uncharged non-dissipative fluid the first non trivial correction appears at second order, which is the one we are interested in. Therefore in this particular case we can use the frame-invariant formulation to simplify the computation despite working at second order.
  
For uncharged fluids the frame invariant combinations that contain the information about the transport coefficients are a transverse traceless two tensor and a scalar:
  \begin{equation}\label{frameinv}
  \begin{split}
  &C_1^{\mu\nu} = P^{\mu \alpha}P^{\nu\beta} \Pi_{\alpha\beta} -\frac{1}{3}\, P^{\mu\nu}\, P^{\alpha\beta}\, \Pi_{\alpha\beta}\,, \\
  &C_2 = \frac{P^{\mu\nu}\Pi_{\mu\nu}}{3} -\frac{s}{T}\left(\frac{dT}{ds}\right)u_\mu u_\nu\Pi^{\mu\nu}\,.
  \end{split}
  \end{equation}
We compute $C_1^{\mu\nu}$ and $C_2$ using the stress tensors computed in  \sec{sec:form} and \sec{sec:entropy} and  compare the two answers. 
 
 Given that stress tensor derived in section \ref{sec:entropy} contains $7$ arbitrary functions and that derived in \eqref{2ndstress} has $5$ functions, we expect the frame invariant combinations to match provided we express the $7$ transport parameters $\{A_3, B_1,B_2, B_3,B_4, \lambda_0, \lambda_2\}$ in terms of the parameters $K_i$ (for $i = 1, \ldots, 5$) appearing in our effective action analysis of \sec{sec:form}. A somewhat tedious algebra\footnote{We need to use the zeroth order equations of motion in \eqref{stressstep} to extract the correct tensor structures; in  particular we make use use of $\nabla_\mu s =s\, \Theta \, u_\mu - T\,\frac{ds}{dT}\, {\mathfrak a}_\mu$.}
   reveals that 
\begin{align}\label{identif}
  &A_3 = -\frac{K_5}{T}, \qquad B_1 = \frac{K_2 - K_1}{2 \,T}, \qquad B_3 = - \frac{K_1 + K_2}{2\, T} 
 \nonumber \\
  & B_2 =- \frac{K_1 + K_2}{6\, T} + \frac{s^2}{T^2}\; \frac{dT}{ds}\; \frac{d K_5}{ds}
 - \frac{s^4}{2\, T}\;  K_3
  \nonumber \\
&B_4 = \frac{K_4}{2\, T} 
- \frac{1}{T^2} \; \frac{dT}{ds}\; \frac{d K_5}{ds}
\nonumber \\
&\lambda_0 = \frac{s}{T}\;\frac{dK_5}{ds} - \frac{2}{3}\; \frac{dK_5}{dT}  + \frac{1}{T} \left(s\; \frac{dK_2}{ds} - K_2+ s\, \frac{dK_1}{ds} - K_1\right)
\nonumber \\
&\lambda_2 =2\, \frac{K_2+ K_1}{T} 
\end{align}
Curiously $\lambda_2$ which does not enter into the entropy current analysis is very simply related to $B_3$; from the above $\lambda_2 = -4\, B_3$ for non-dissipative fluids arising from an effective action.

In a generic frame the expression for the full entropy current is given in \eqref{ecurrentG}. In fact in this expression one can show that $s \, u^\mu - \frac{1}{T}\, u_\nu\, \Pi^{\mu\nu}$ is frame invariant, thereby demanding that $J^\mu_{cor}$ is insensitive to field redefinitions. Given that the analysis of \sec{sec:form} is in the entropy frame, one is led to the following conclusion: $\frac{1}{T}\, u_\nu\, \Pi^{\mu\nu}$ computed using \eqref{2ndstress}  should be equal to $J^\mu_{cor}$ of \eqref{entcurgen} once we substitute the identifications as given in \eqref{identif}. This algebraic check works out as expected and along the way we get a useful bonus. One is now able to fix the two remaining parameters of the entropy current $A_1$ and $A_2$ which by virtue their vanishing divergence remain unconstrained. One finds
\begin{equation}
A_1 = -\frac{1}{T}\; \frac{dK_5}{dT}\,, \qquad A_2 =\frac{K_2 - K_1}{T^2}
\label{}
\end{equation}	
It would be interesting  to understand why these relations are forced upon the fluids that arise out of the effective action constructed using $\text{Sdiff}({\cal M}_\phi)$ invariance.

Finally, note that we can fix the functions $K_i$ to ensure that the fluid is conformal, by using the constraints on the functions entering the entropy current analysis and \eqref{identif}.  We find:\footnote{Notice that we are not a-priori demanding scale invariance of our action, but rather using the identification between the action and entropy current analyses \eqref{identif} to constrain the functions $K_i$ appearing in the action \eqref{2ndaction}. It is easy  to convince oneself that the counting works out correctly by noting that the conformally covariant  scalars are three in number: these are built out of $\sigma_{\mu\nu}\, \sigma^{\mu\nu}$ \, $\omega_{\mu\nu}\, \omega^{\mu\nu}$ and a specific linear combination of $\{ \Theta^2, {\mathfrak a}^2, R\}$. We have refrained from translating this into the basis we chose to work with preventing a direct comparison immediately; the constraint on $K_i$  \eqref{Kiconf}, should follow directly from this translation. } 
\begin{align}
&K_1  = -(b_1 + b_3)\, s^\frac{2}{3} \,,\quad K_2 = (b_1 + b_3)\, s^\frac{2}{3} \,,
\nonumber \\
& K_3 = -\frac{2}{3}\, b_3\, s^{-\frac{10}{3}} \,, \quad \;K_4 = -\frac{2}{3}\, a_3\, s^{-\frac{4}{3}} \,, \qquad K_5 = -a_3\, s^\frac{2}{3}\,,
\label{Kiconf}
\end{align}	
accounts for all the constraints. Since now $\ell_2$ which determines $\lambda_2$ is fixed in terms of $b_3$ from \eqref{identif}, we learn that the conformal fluids which arise from an action are required to satisfy one additional relation.  Using the parameterization \eqref{ABpars} the transport coefficients for conformal fluids arising from the action principle can be encapsulated as:
\begin{align}
& \tau = 3\, \lambda_0 = 2\,(a_3+b_3)\, T \,, \qquad \kappa_2 = 2\,\kappa_1 = \lambda_1 = 2\, a_3\, T\,, 
\nonumber \\
& \lambda_2 = -4\,b_3\,T \,,\qquad\qquad  \lambda_3 = 2\,(a_3\, - 2\, b_1) \, T
\label{cfnd}
\end{align}	
Thus, the action formalism only allows us to explore a three parameter family of  non-dissipative conformal fluids.

%______________________________________________
\section{Conclusion}
\label{sec:conclude}
%______________________________________________

Motivated by a need to understand the fundamental constraints on an autonomous theory of hydrodynamics, we have in this note explore neutral non-dissipative fluids. In particular, we have examined the conditions on such fluids (neutral) arising from demanding the existence of a divergence-free entropy current as well as from an action principle. While both these analyses are consistent with each other, we find a larger parameter family of non-dissipative fluids from an entropy current analysis.

The entropy current formalism determines $\{2+13\}$ out of a total $\{2 +15\}$ transport coefficients of the first and second order stress tensor in terms of the 5 free parameters appearing in the most general zero divergence entropy current  (thereby predicting 8 linear relations among 13 of  second order transport coefficients). Two second order transport coefficients do not enter into the entropy current analysis: these are $\lambda_2$ which multiplies the (shear) $\times$ (vorticity) contribution and a linear combination $\lambda_0 - \xi_2$ which multiplies a trace-free combination of expansion and shear. All in all we have a seven parameter family of second order non-dissipative fluids which are consistent with absence of entropy production for arbitrary flows.

On the other hand, in the effective action action approach the existence of a zero divergence entropy current is ensured by demanding the reparameterization invariance in field space $\text{Sdiff}({\cal M}_\phi)$. We have found a five parameter family of effective actions at second order, which lead thence to a stress tensor where all the $\{2+15\}$ transport coefficients are determined in terms of these $5$ parameters. The resulting stress tensor is of course a special case of the one predicted by the entropy analysis and upon suitable identification of variables one is also led to fixing the free transport coefficients $\lambda_2$ and $\lambda_0 - \xi_2$ of the latter construction.

The results we have obtained are given explicitly for both conformal and non-conformal fluids in the Landau frame and in the entropy frame. The former is convenient for comparison with standard results in the literature, but as been noted elsewhere it is easy to pass between the frames by a redefinition of fluid-dynamical fields.

It is interesting to ask about the physical interpretation of the transport coefficients in the absence of dissipation. First of all, note that even though there are no non-dissipative first order transport coefficients for neutral fluids, the second order terms do affect transport. The simplest manifestation of this is in the dispersion relation about equilibrium. Consider a fluid on Minkowski spacetime: the equilibrium solution is constant temperature and velocity field (pointing along the timelike Killing field). Linear fluctuations about this are characterized by plane waves and it is easy using the explicit form of the stress tensor \eqref{stresderi1} and \eqref{fconst1} to work out the effects on the dispersion. We find that the sound mode gets corrected by 
contributions from $\tau$ and $\zeta_1$:\footnote{All other terms in \eqref{stresderi1} involve terms which are quadratic order in the linearised analysis. For $u^\mu = u^\mu_{(0)} + \delta u^\mu\, e^{-i\, \omega \, t + i\, {\bf k}\cdot {\bf x} }$ and $ T= T_{(0)} + \delta T\, e^{-i\, \omega \, t + i\, {\bf k}\cdot {\bf x} } $ with $\delta T$ and $\delta u^\mu$ of order $\varepsilon$ we note that only terms multiplying $\tau$ and $\zeta_1$ are ${\cal O}(\varepsilon)$.} explicitly, 
\begin{equation}
\omega  = \pm \, v_s\, k \left[1 + k^2\; \frac{T_0}{\varepsilon_0 + P_0} \left(\frac{\tau(T_0)}{3} + \frac{\zeta_1(T_0)}{2}\right)\right]
\label{}
\end{equation}	
with $v_s = \sqrt{dP/d\varepsilon}$ as usual and the subscript $0$ stands for the equilibrium value. So while the higher order  transport do not change the speed of sound (which they cannot since the latter is thermodynamic), they do affect the sub-leading parts of the dispersion.\footnote{We are assuming arbitrary values of $\tau$ and $\zeta_1$ here; it is a simple matter to replace these by the specific parameterisations encountered earlier.} While 
the other transport do not enter into the dispersion in flat space, they do affect other flows: for conformal fluids this has been described in \cite{Baier:2007ix, Bhattacharyya:2008jc}. Thus by judiciously engineering flows which are sensitive to various combinations of shear, expansion, etc., one can read off all the transport coefficients.

The main outstanding question of our analysis is why the effective action approach gives a smaller parameter family of second order non-dissipative fluids than that predicted by demanding the presence of a zero divergence entropy current. A-priori one can think of two distinct reasons for this mismatch:
\begin{itemize}
\item[(i)] We have not identified all possible dimension two scalar operators built out of the $\phi^I$, the basic variables entering into the effective action. As we discussed in \sec{sec:form} we worked out the independent structures that preserved the volume form on the $\phi^I$ manifold ${\cal M}_\phi$ using structures built out of the current $J^\mu$ and its derivatives. It is plausible (though we think unlikely) that there are other admissible structures.
\item[(ii)] More intriguing (assuming we have identified all possible contributions to the action) is the possibility that there are indeed further constraints on fluid dynamics coming from the existence of an effective action, which are not simply captured in terms of symmetry constraints on the stress tensor and demanding the local form of the second law (or rather zero divergence for the case of non-dissipative fluids). If this were the case, then we would have an interesting window of opportunity of using these non-dissipative fluids to learn about the potential constraints on hydrodynamic expansions. Perhaps out results hint at additional microscopic constraints on the transport coefficients $\lambda_2$ and $\lambda_0 -\xi_2$?
\end{itemize}

The  mismatch in the parameter count is the salient feature of our analysis and its shows up at second order for neutral fluids. We would be remiss to not point out that the effective action approach can be seen to give a restricted class of fluids when we allow for parity-violation already at first order.  The authors of \cite{Nicolis:2011ey} have previously used this approach to describe Hall viscosity (a new transport coefficient) in three dimensions. This parameter multiplies the parity-odd $\epsilon^{\alpha\rho(\mu} \, u_\alpha \, \sigma^{\nu)}_{\;\rho}$ term in the stress tensor. One can show that the coefficient here is entropy preserving. In fact it does not enter into the entropy current analysis (for pretty much the same reason as $\sigma^{(\mu}_{\;\alpha} \, \omega^{\nu)\alpha}$ in our analysis of \sec{sec:entropy}). Curiously this term is again not reproduced from the canonical local action one writes down. Moreover,  the entropy current following the recent analysis of \cite{Jensen:2011xb}  (who also included a conserved global $U(1)$  current) reveals a two parameter family of parity-odd dissipation less fluids, while the effective action predicts a single parameter (see \App{s:parityodd} for details). 
Likewise in the analysis of \cite{Dubovsky:2011sk} who explored the possibility of describing anomalous charged fluid transport in $d=2$ using this effective action approach, one obtains a one-parameter family of anomalous charged fluids at first order in gradients. On the other hand entropy considerations \cite{Loganayagam:2011mu,Loganayagam:2012pz} as well as more recent analysis using equilibrium partition functions \cite{Jain:2012rh} clearly show a two parameter family of such fluids (only one of the parameters is related to the global current anomaly).

It would be interesting to extend this analysis to charged fluids, superfluids and fluids with anomaly (in higher dimensions) to see whether we obtain more constraints from demanding an effective action's existence.

At the end of the day if we believe the implications of the effective action formalism, we should be able to derive the constraints on transport directly from microscopic theory. A useful avenue for exploration is perhaps analysis of stress tensor correlation functions. For instance as shown in \cite{Moore:2010bu}  (see also \cite{Moore:2012tc} for non-conformal fluids)  one can derive Kubo formulae for the second order transport coefficients and in particular for $\lambda_2$. One should examine whether the condition of vanishing $\eta$ forces constraints on the particular three-point function determining $\lambda_2$.

While still on the topic of fluid dynamics,  in \cite{Banerjee:2012iz, Jensen:2012jh} the authors have written a partition function or a generating functional for stress tensor evaluated on the most general time independent fluid flow on any arbitrary but static background. One can check using the action \eqref{2ndaction}, that there is a three-paramater family of equilibrium actions (essentially $\sigma_{\mu\nu} = \Theta =0$ for stationarity), which agrees with the counting based on sources that be turned on maintaining time independence. 
It would be interesting to explore the connection between these two formalisms. In particular it would be nice to know how one can determine the partition function describing the time independent situation from the action written for a time dependent but strictly non-dissipative fluid flows.

 The bulk of this paper has been concerned with analysis of fluid dynamics in its own right as an effective action. 
 We have not so far touched upon the issue of whether non-dissipative fluids are physical, nor have we mentioned any connections to holography.  Now is the time to remedy these lacunae in our discussion.
 
 As we remarked in \sec{sec:intro} the fact that on-dissipative fluids demand $\eta  = \zeta=0$ makes one suspect that this type of hydrodynamics cannot arise from a sensible unitary quantum field theory. Indeed while $\zeta = 0$ is realized in conformal fluids, one expects on general grounds the viscosities to be bounded from below from general arguments. For instance using the uncertainty principle one can generically argue that $\eta \geq \alpha\, \frac{\hbar}{k_B}\, s$ where for a change we have explicitly indicated the fundamental constants. There is an undetermined constant $\alpha \sim {\cal O}(1)$ in the above analysis. Indeed it has been speculated following the seminal work of \cite{Kovtun:2004de} that there is a lower bound on shear viscosity; the status of this bound has been the subject of much scrutiny and the current wisdom is that $\frac{\eta}{s} \geq \frac{\gamma}{4\,\pi}$ where $\gamma$ is a fraction which seems to depend on the details of microscopic theory (see \cite{Sinha:2009ev} for a recent review of the bound). Likewise, while $\zeta =0$ is forced upon one by scale invariance, in non-scale invariance theories it is also conjectured that there is a lower bound on $\zeta$, viz., $\zeta \geq 2\,\eta  \left(\frac{1}{d-1} - c_s^2\right)$ with $c_s$ being the speed of sound \cite{Buchel:2007mf} (this bound is much less explored). Based on these arguments one might indeed be tempted to take the viewpoint that non-dissipative fluids of the type discussed herein are unphysical. This is supported for instance by holographic computations of shear viscosity in higher curvature gravity theories; \cite{Brigante:2007nu,Brigante:2008gz} studied $\eta/s$ in Gauss-Bonet-AdS gravity  and argued that while one could use the Gauss-Bonet coupling to lower $\eta$ (in fact all the way to zero) demanding sensible causal properties of the bulk (most likely related to unitarity in the field theory) results in a lower bound for $\eta$ away from zero. It would be interesting to explore this example in some detail to understand whether the fine tuned Gauss-Bonet theory with $\eta =0$ has non-trivial second order transport coefficients which refrain from entropy production.
 
 We should also point out absence of dissipation could imply that generic flows suffer from a turbulent instability, with its consequent energy cascade. The absence of dissipation would be relevant in that the system will not be able to exit the cascade gracefully (as happens in physical systems with dissipation). However, it is not clear to us what the effect of  the non-linear effects engendered by the second order coefficients is on the inertial range of the flow (assuming it is driven turbulent).  This is an interesting question, which deserves to be explored further. 
 Another potential consequence of our idealized non-dissipative fluid is that random thermal fluctuations could drive the system away from equilibrium, cf., \cite{Kovtun:2012rj} for a comprehensive discussion. Given that one has an action formalism, one should be able to compute these effects using standard techniques; we hope to return to this issue in the future. 
 
  Finally, one of the  reasons to take this explorations seriously is the potential it has to teach us about honest dissipative fluids. One might ask how could one deform away from the situation where entropy is conserved to one where the local form of second law is valid in standard form. A clue comes from the origin of the identity $\nabla_\mu J^\mu =0$; this is forced upon us by the fact that we demand invariance under volume preserving diffeomorphisms on $\phi$ space. Suppose we were to relax this condition to just demanding diffeomorphism invariance -- this is a sensible requirement to impose since we are still allowed to relabel fluid elements arbitrarily. Can one then write down an effective action for dissipative fluids? The tentative answer coming from the analysis of  \cite{Nickel:2010pr} is yes: by using holographic intuition for the renormalization group these authors have constructed a leading order effective action which seems to capture the effects of shear viscosity. Indeed as they point out in the holographic context the fields $\phi^I$ which we view abstractly as labels for the fluid, pick up a fascinating geometric meaning. Since the fluid/gravity correspondence \cite{Bhattacharyya:2008jc} asserts that arbitrary fluid configurations of a field theory with a holographic dual is given by a  black hole solution in an asymptotically AdS spacetime with a regular event horizon,\footnote{This assertion extends beyond AdS to encompass long-wavelength world-volume fluctuations of black branes as exemplified in the blackfold approach \cite{Camps:2012hw}.} we can parameterize our solution in terms of the spatial geometry of the horizon which is then Lie transported along the future horizon generator. The natural coordinates on these spatial sections are nothing but the $\phi^I$: in equilibrium (or indeed in the absence of dissipation) this follows from the analysis of gravitational entropy current of \cite{Bhattacharyya:2008xc}. It would be interesting to explore these relations in greater detail and we hope that future studies enables one chart out a clear autonomous theory of hydrodynamics.
  
\bigskip  
\noindent
{\em Note added in v3:} From \eqref{cfnd} (having fixed a typo) we see that the transport coefficients $\tau$, $\lambda_1$ and $\lambda_2$ obey a linear relation
  \begin{align}
  \tau = \lambda_1 - \frac{1}{2} \,\lambda_2
  \label{}
  \end{align}
  This is not specific to conformal fluids, this also holds more generally as can be seen from the expressions for these quantities in \eqref{lam13sol} and \eqref{identif}. Curiously this is the same relation that has been argued to hold universally for holographic theories \cite{Haack:2008xx}. We thank  R.~Loganayagam and A.~Yarom for discussions which led to this observation.
 
%~~~~~~~~~~~~~~~~~~~~~~~~~~~~~~~~~~~~~~~~~~~~~~
\acknowledgments 
%~~~~~~~~~~~~~~~~~~~~~~~~~~~~~~~~~~~~~~~~~~~~~~
It is a pleasure to thank Veronika Hubeny, Sachin Jain, Hong Liu, Arnab Rudra, Ashoke Sen, Tarun Sharma,  Tadashi Takayanagi, Mauricio Andres Romo,  Yun-Long Zhang, and especially R. Loganayagam and Shiraz Minwalla for numerous enlightening discussions on topics related to hydrodynamics, entropy and effective actions. 

We would like to thank the Isaac Newton Institute for hospitality during the program ``Mathematics and Applications of Branes in String and M-theory" where this work was initiated. SB and MR would also like to thank TIFR for providing a stimulating environment during the course of this project. Finally JB and MR would like to thank the Yukawa Institute for Theoretical Physics during the YKIS2012 symposium for their hospitality during the concluding stages of this project. The work of JB was supported by World Premier International Research Center Initiative (WPI Initiative), MEXT, Japan. 
MR is supported in part by the  STFC Consolidated Grant ST/J000426/1.
 
\appendix

%~~~~~~~~~~~~~~~~~~~~~~~~~~~~~~~~~~~~~~~~~~~~~~~
\section{Some useful results}
\label{s:conventions}
%~~~~~~~~~~~~~~~~~~~~~~~~~~~~~~~~~~~~~~~~~~~~~~ 
 
\paragraph{Conventions:} We collect some useful formulae in this appendix. First with regard to conventions:  in the text we use the standard (anti-)symmetrisation convention i.e., 
$A^{(\mu\nu)} \equiv \frac{1}{2}\left(A^{\mu\nu} + A^{\nu\mu}\right)$.
We define the curvature tensors as:
\begin{equation}
{R^\rho}_{\alpha\beta\nu} = \partial_\beta \Gamma^\rho_{\alpha\nu} -
\partial_\nu \Gamma^\rho_{\alpha\beta} 
+ \Gamma^\lambda_{\alpha\nu}\Gamma^\rho_{\lambda\beta}
-\Gamma^\lambda_{\alpha\beta}\Gamma^\rho_{\lambda\nu} \,, \qquad R_{\mu\nu} = {R^\rho}_{\mu\rho\nu}
\label{}
\end{equation}	

\paragraph{Useful variational formulae:} The reader will find the following useful to derive \eqref{2ndstress}: the variation of $J^\mu, ~~ u^\mu, ~~s $ and $\Gamma^\mu_{\theta\phi}$,( the connection) with respect to the metric are given by 
\begin{equation}\label{idex2}
\begin{split}
\delta J^\mu &= -\left(\frac{J^\mu}{2}\right) g^{\alpha\beta}~\delta g_{\alpha\beta}\\
\delta u^\mu &= \left(\frac{u^\mu}{2}\right) u^{\alpha} u^{\beta}~\delta g_{\alpha\beta}\\
\delta s &= -\left(\frac{s}{2}\right) P^{\alpha\beta}~\delta g_{\alpha\beta}\\
\delta\Gamma^\mu_{\theta\phi} &= -\left(\frac{\delta g_{\alpha\beta}}{2}\right)\left[g^{\mu\alpha}\left(\delta^\beta_\theta\nabla_\phi + \delta_\phi^\beta\nabla_\theta\right) - \delta^\alpha_\theta \delta^\beta_\phi \nabla^\mu\right]
\end{split}
\end{equation}

\paragraph{Eliminating the velocity projected curvature scalar:} To show that $R_{\mu\nu} \,u^\mu\, u^\nu$ can be eliminated on-shell note that
\begin{equation}\label{totalderi}
\begin{split}
&~~\int {\sqrt g}~K(s) R_{\mu\nu} u^\mu u^\nu = -\int {\sqrt g}~ K(s)\, [\nabla_\nu,\nabla_\rho]\, u^\rho u^\nu\\
=&~~\int {\sqrt g}~K(s) \left[ u^\nu \, \nabla_\rho \nabla_\nu u^\rho  + u_\alpha\nabla^\alpha \left(u^\beta\nabla_\beta s\right)\right]\\
=&-\int {\sqrt g}\left\{K(s)(\nabla_\rho u^\nu)(\nabla_\nu u^\rho)+ [(u.\nabla)u^\rho] \partial_\rho K(s) + (u.\partial s)\nabla_\mu\left[K(s) u^\mu\right]\right\}\\
=&-\int {\sqrt g}\left\{K(s)(\nabla_\rho u^\nu)(\nabla_\nu u^\rho)-T\frac{dK}{dT}[(u.\nabla)u^\rho] ^2 + \frac{d}{ds}\left(\frac{K(s)}{s}\right) (u.\partial s)^2\right\}
\end{split}
\end{equation}

\paragraph{Simplifications of the stress tensor derived in \sec{sec:form}:}
Some of the expressions appearing in the stress tensor derived from the action can be simplified when we use the decomposition \eqref{pudecomp} and also the zeroth order equations of motion. The latter can be cast in a useful form which reads:
\begin{equation}
\nabla_\mu s =s\, \Theta \, u_\mu - T\, \frac{ds}{dT}\, {\mathfrak a}_\mu
\label{eomT1}
\end{equation}	

For reference we record that the combinations appearing in \eqref{stena} and \eqref{stenb} can be simplified into the form:
\begin{align}
&-\nabla_\alpha\left[u^{(\mu} P^{\alpha\beta}K_1 \nabla_\beta u^{\nu)}\right] 
- \nabla_\alpha\left[u^\alpha \,P^{\beta(\mu} K_1 \nabla_\beta u^{\nu)}\right]
+ \nabla_\alpha\left[K_1 u^{(\mu} P^{\nu)\beta}\nabla_\beta u^\alpha\right]  \nonumber \\
& = -\nabla_\alpha\left(K_1\left[u^\alpha\, \sigma^{\mu\nu} -2\, u^{(\mu}\,\omega^{\nu)\alpha}  + \frac{\Theta}{d-1}\,  P^{\mu\nu}\, u^\alpha\right]\right)
\label{}
\end{align}	
and 
\begin{align}
& -\nabla_\alpha\left[K_2 u^{(\nu} \nabla^{\mu)} u^\alpha\right] 
- \nabla_\alpha\left[K_2 u^\alpha \nabla^{(\mu} u^{\nu)}\right] 
+ \nabla_\alpha\left[K_2 u^{(\nu} \nabla^\alpha u^{\mu)}\right]  
\nonumber \\
&= -\nabla_\alpha \left(K_2\left[u^\alpha\, \sigma^{\mu\nu} + 2\, u^{(\mu}\, \omega^{\nu)\alpha} - a^\alpha\, u^\mu\,u^\nu + \frac{\Theta}{d-1}\, P^{\mu\nu}\, u^\alpha \right]\right)
\label{}
\end{align}
This is sufficient to read off some of the transport coefficients, especially $\tau$ directly.

%~~~~~~~~~~~~~~~~~~~~~~~~~~~~~~~~~~~~~~~~~~~~~~~
\section{Parity-odd neutral non-dissipative fluids in $d=3$}
\label{s:parityodd}
%~~~~~~~~~~~~~~~~~~~~~~~~~~~~~~~~~~~~~~~~~~~~~~ 
%
In this appendix we will consider parity-odd neutral fluids in $2+1$ dimensions at first order in derivative expansion. As we have seen in \sec{sec:form} there are no non-dissipative terms 
at first order in hydrodynamics in $3+1$ dimensions. This fact is also true in $2+1$ dimension, if we preserve parity. However if we consider parity violating effects then it is possible 
to write a first order term in the action using our basic variables $\phi^I$, which is manifestly invariant under reparameterizations of the the fields \cite{Nicolis:2011ey}.  The desired  term has the form 
\begin{equation}\label{poaction}
 S_1^{\text{(parity-odd)}} = \int d^3x \,\sqrt{-g} \,  \varsigma(s) \, \epsilon^{\mu \nu \lambda} \, u_{\mu} \, \nabla_{\nu} u_{\lambda} = \int d^3x \, \sqrt{-g} \, \varsigma(s) \, \Omega 
\end{equation}
where $\varsigma(s)$ is some arbitrary function of the entropy density which will ultimately determine the transport coefficients appearing in the stress tensor. 
Here
\begin{equation}\label{omegadef}
 \Omega \equiv   \epsilon^{\mu \nu \lambda} u_{\mu} \nabla_{\nu} u_{\lambda}
\end{equation}
is the vorticity which is a scalar in $2+1$ dimensions. Our conventions are that  $ \epsilon^{\mu \nu \lambda} = \frac{1}{\sqrt{-g}} \, {\tilde \epsilon}^{\mu \nu \lambda} $  where ${\tilde \epsilon}^{\mu\nu\rho}$ is the flat space Levi-Civita tensor, with orientation ${\tilde	\epsilon}^{012} = +1$. 

The first order corrections to the stress tensor that follows from the action \eqref{poaction} is given by 
\begin{equation}\label{poPi}
\begin{split}
 \Pi^{\mu \nu}_{\text{(parity-odd)}} = &- \frac{s}{2} \,\varsigma'(s)\, \Omega \,  P^{\mu \nu} +  \varsigma(s) \, \Omega \, u^{\mu} u^{\nu}
\\ &+ 2 \,\varsigma(s) \,  \epsilon^{(\mu \rho \lambda}  u^{\nu)}  \nabla_{\rho} u_{\lambda}
+\varsigma'(s) \,  \epsilon^{(\mu \rho \lambda} \, u_{\lambda}\; \nabla_\rho s \;    u^{\nu)} .
\end{split}
\end{equation}
Note that there is no frame invariant genuine tensor in this expression of the stress tensor. In other words if we take $  \Pi^{\mu \nu}_{\text{(parity-odd)}} $ from \eqref{poPi}
and evaluate $C^{\mu \nu}_1$ using \eqref{frameinv} suitably modified for $2+1$ dimensions, we find that it evaluates to zero. 
There is however, a frame invariant scalar data in \eqref{poPi}, since for this correction to stress tensor $C_2$ in \eqref{frameinv} evaluates to 
\begin{equation}\label{pofidata}
 C_2^{\text{(parity-odd)}} = -\left( \frac{s}{2} \varsigma'(s) +   \frac{s}{T} \frac{d T}{d s} \varsigma(s) \right) \Omega
\end{equation}

Now let us consider the most general parity-odd corrections to the stress tensor and canonical entropy current, purely based on symmetry grounds. We will then  explore the restrictions imposed on these corrections from the demand that the entropy current be divergence free. parity-odd fluid dynamics was recently examined in \cite{Jensen:2011xb}
quite generally (including the presence of a global $U(1)$ charge). In fact, we can directly take over their analysis by setting the charges to zero; the parity-odd terms which are allowed in the neutral fluid are naturally entropy conserving. Here we reproduce their result by restricting ourselves  to the case of  neutral fluids.

As in \S \ref{sec:entropy} we now work in the Landau frame for this calculation. 
At first order the possible parity-odd corrections for the neutral fluid is given by \cite{Jensen:2011xb}\footnote{ Note that the $\Omega$ defined in \cite{Jensen:2011xb} differs from that defined in \eqref{omegadef} by a minus sign.}
\begin{equation}\label{poST}
 \Pi^{\mu \nu}_{\text{(parity-odd)}} = - \tilde{\eta} \, \tilde \sigma^{\mu \nu} + \tilde \chi_\Omega \, \Omega \, P^{\mu \nu},
\end{equation}
where 
\begin{equation}\label{tsigdef}
 \tilde \sigma^{\mu \nu} = \epsilon^{(\mu \alpha \beta} u_\alpha \sigma_{\ \beta}^{\nu)} 
% ; \quad \Omega = \epsilon^{\mu \nu \lambda} u_{\mu} \partial_{\nu} u_{\lambda}.
\end{equation}
is the parity-odd dual of the shear tensor; the corresponding transport coefficient $\tilde{\eta}$ is referred to as the Hall viscosity.  Note that we do not consider any parity-even corrections to the stress tensor because we know that the shear and bulk viscosities have to vanish to ensure dissipation-free behaviour of our fluid.

The first order equations of motion \eqref{eomT1} allow us to eliminate the gradient of the entropy or equivalently the temperature in terms of velocity gradients. As a consequence of this equation of motion there is only one transverse pseudo-vector and one pseudo-scalar at first order which are respectively given by 
\begin{equation}\label{po1ord}
 U^{\mu} = \epsilon^{\mu \nu \rho} u_{\nu} \, u^\alpha\, \nabla_\alpha  u_{\rho}\,, \quad \Omega = \epsilon^{\mu \nu \rho} u_{\mu} \partial_{\nu} u_{\rho}. 
\end{equation}
Therefore at first order in gradients there are only two possible on shell linearly independent vectors; we choose to write these two vectors as
\begin{equation}\label{Vdef}
 \begin{split}
  V_1^{\mu} &= \epsilon^{\mu \nu \rho} u_{\nu} \nabla_{\rho} T=-T U^{\mu}\\
  V_2^{\mu} &= \epsilon^{\mu \nu \rho} \nabla_{\nu} u_{\rho} = -U^{\mu} - u^{\mu} \,\Omega.
 \end{split}
\end{equation}
Using these independent vectors the most general entropy current at first order involving parity-odd terms is:
\begin{equation}\label{entcurpo}
 J^{\mu}_{(s)} = J^{\mu}_{\text{can}} + \alpha_1 \, V_1^{\mu} + \alpha_2 \, V_2^{\mu} 
\end{equation}
We could have considered adding parity-even corrections to the entropy current in \eqref{entcurpo}. These however  that do not affect the parity-odd  analysis; moreover as described in \cite{Jensen:2011xb} the coefficients of such parity-even corrections would be set to zero when we demand that there is no local entropy production. 

Now using the $2+1$ dimensional version of \eqref{prev1} we can directly evaluate the divergence of the canonical part of the entropy current in the parity-odd sector, we find
\begin{equation}\label{poCentcurdiv}
\begin{split}
 \nabla_{\mu} J^{\mu}_{\text{can}} &= - \frac{1}{T} \left( \sigma_{\mu \nu} \Pi^{\mu \nu} + \frac{\Theta}{2} P_{\mu \nu} \Pi^{\mu \nu} \right)
 = -\frac{1}{T} \,\tilde \chi_{\Omega} \, \Theta \, \Omega.
\end{split}
\end{equation}
The divergence of the remaining two terms are given by
\begin{equation}\label{poentdiv2}
\begin{split}
 \nabla_{\mu} \big( \alpha_1 \, V_1^{\mu} + \alpha_2 \, V_2^{\mu} \big) &= \frac{d \alpha_1}{d s} \, V_1^{\mu} \, \nabla_{\mu} s + \frac{d \alpha_2}{d s} \,V_2^{\mu} \, \nabla_{\mu} s
+ \alpha_1 \, \nabla_{\mu} V_1^{\mu} + \alpha_2 \, \nabla_{\mu} V_2^{\mu} \\
&= s\, \left(  \frac{d \alpha_2}{d s} + \frac{d T}{d s} \alpha_1 \right)  \Theta \Omega 
\end{split}
\end{equation}
Thus divergence of the full entropy current is given by 
\begin{equation}\label{entcurdivpo}
 \nabla_{\mu} J^{\mu}_{(s)} = \bigg[-\frac{1}{T} \,\tilde \chi_{\Omega} 
+s \left(  \frac{d \alpha_2}{d s} + \frac{d T}{d s} \alpha_1 \right)  \bigg]  \Theta \, \Omega 
\end{equation}
This implies that for zero divergence of the entropy current we must have ${\tilde \chi}_\Omega$ fixed in terms of the coefficients appearing in the entropy current
\begin{equation}\label{chiOm}
 \tilde \chi_{\Omega} = T\,  \left(  \frac{d \alpha_2}{d s} + \frac{d T}{d s} \alpha_1 \right) .
\end{equation}
Since the parity-odd shear tensor ${\tilde \sigma}_{\mu\nu}$  is passive in the entropy analysis, it follows ${\tilde \eta}$ remains completely unconstrained. 

Therefore we see that the existence of a divergence-free of the entropy current be zero fails to constrain either  of the two parity-odd first order  transport coefficients in \eqref{poST}. One of these coefficients, the Hall viscosity, does not enter the divergence of the entropy current. The other is determined in terms of the possible corrections to the canonical form of the entropy current which are arbitrary at this order.

There are two pieces of frame invariant data in \eqref{poST}, that is both $C_1^{\mu \nu}$ and $C_2$ are non-zero when we plug in \eqref{poST}
into \eqref{frameinv} and they are given by
\begin{equation}\label{fientcurpo}
 \begin{split}
  C_1^{\mu \nu} &= - \tilde{\eta} \, \tilde \sigma^{\mu \nu} \\
  C_2 &= \tilde \chi_\Omega \, \Omega.
 \end{split}
\end{equation}
Comparing this result with the frame invariant data present in the stress tensor following from the action \eqref{poaction}, we conclude that 
through the action we are only able to capture the effect of the vorticity ($\Omega P_{\mu \nu}$) term and $\tilde \chi_{\Omega}$ is determined in terms of the 
function $\varsigma(s)$ in the action through the following relation 
\begin{equation}
 \tilde \chi_{\Omega} =  -\left( \frac{s}{2} \varsigma'(s) +   \frac{s}{T} \frac{d T}{d s} \varsigma(s) \right) 
\end{equation}
The Hall viscosity naively seems to be undetermined from the action formalism. 

In passing we note that it may be possible to make modifications to the stress tensor \eqref{poPi}  as has been argued on physical grounds  in \cite{Nicolis:2011ey}; the  modified stress tensor then can captures the effect of the Hall viscosity. We do not see such modifications directly from at the level of the action. In any event without any such modifications or improvements to the  stress tensor, the one which directly follows from the action \eqref{poaction} captures only transport corresponding to vorticity $\Omega$.

 %%%%%%%%%%%%%%%%%%%%%%%%%%%%%%%%%%%%%%%%%%%%%%%%
% 
%\bibliographystyle{JHEP}
%\bibliography{non-dissipative}
\providecommand{\href}[2]{#2}\begingroup\raggedright\endgroup

\end{document}